\documentclass{cpcauth}	
\usepackage{epsfig}
\begin{document}	
\begin{frontmatter}
\title{Classical and ab initio molecular dynamic simulation of an amorphous silica surface}
\author{C. Mischler$^1$, W. Kob$^2$ and K. Binder$^1$}
\address{$^1$Institut f\"ur Physik, Johannes Gutenberg-Universit\"at
\\55099 Mainz; Germany\\
$^2$Laboratoire des Verres, Universit\'e Montpellier II \\
34095 Montpellier; France}

\begin{abstract}
We present the results of a classical molecular dynamic simulation
as well as of an {\it ab initio} molecular dynamic simulation of an
amorphous silica surface. In the case of the classical simulation we use
the potential proposed by van Beest {\it et al.} (BKS) whereas the {\it
ab initio} simulation is done with a Car-Parrinello method (CPMD). We
find that the surfaces generated by BKS have a higher concentration of
defects (e.g. concentration of two-membered rings) than those generated
with CPMD. In addition also the distribution functions of the angles and
of the distances are different for the short rings. Hence
we conclude that whereas the BKS potential is able to reproduce correctly
the surface on the length scale beyond $\approx$5~\AA, it is necessary
to use an {\it ab initio} method to predict reliably the structure at
small scales.

\end{abstract}
\begin{keyword} glass surface; silica; molecular dynamics simulations; 
ab initio simulations
\end{keyword}
\end{frontmatter}

\section{Introduction}
Obtaining a good understanding of the structural and dynamical
properties of the surface of amorphous silica is very important for
the manufacture of glass as well as the construction of electronic
devices~\cite{legrand98}. This is the reason why in the past a large
number of experiments have been done to investigate this type of
surface. Since in real experiments it is rather difficult to obtain
reliably details on the structure also quite a few computer simulations
have been done in order to study this system (see~\cite{roder01} and
references therein). Most of these studies have, however, been done by
using effective classical potentials, such as, e.g., the one proposed
some years ago by van Beest, Kramer, and van Santen (BKS)~\cite{beest90}.
Although it has been shown that these type of potentials can reproduce
quite reliably the structure and dynamics of silica in the {\it bulk}~(
see, e.g., \cite{horbach99} and references therein), it is much less
obvious to what extent they are also able to give a correct description
of the properties of silica close to a surface, since the parameters for
these potentials, effective charges, etc., have often been optimized
to reproduce only experimental data for the bulk. One possibility to
avoid this problem with the classical effective potential
is to use {\it ab initio} simulations such as the scheme proposed
by Car and Parrinello~\cite{car85} since in this type of approach an
effective potential between the ions is calculated self consistently on
the fly, i.e. the instantaneous geometry of the ions is always taken
into account. The drawback of this approach is that due to the huge
computational burden only relatively short time scales, a few ps, 
as well as small systems, a few hundred particles, can be
simulated, whereas classical simulations allow to simulate thousands of
particles over several ns.

In the present work we compare the results of a classical simulation of an
amorphous silica surface with the BKS potential with the results obtained
by the Car-Parrinello Molecular Dynamics method (CPMD). The goal is to
check which quantities are reproduced correctly by the BKS potential,
using the results of the CPMD simulation as the reference system.

\section{The BKS potential and the setup of the geometry}

We have first prepared the system using the BKS potential. In this
two-body potential the atoms interact also by means of a Coulomb
potential where the effective charges of a silicon and oxygen atom is
2.4 and $-1.2$, respectively. More details on this potential can be
found in Ref.~\cite{beest90}.

In order to minimize finite size effects as well as surface effects
it is customary to use periodic boundary
conditions (PBC) in all three directions. If one wants to investigate a
free surface, the most straightforward idea is to use a film geometry,
i.e. to have PBC in two directions and to have an infinite free space
above and below the system. Unfortunately it turns out, however, that
from a computational point of view this setup is not very good for
systems with Coulombic interactions (i.e.  such as the one studied
here), since it prevents to make an {\it efficient} use of the Ewald
summation method. Therefore we have adopted the following strategy
which is explained also in Fig.~\ref{fig1} (see also \cite{feuston89,ceresoli00}):
\begin{figure}[ht]
\begin{center}
\unitlength1mm
\begin{picture}(0,100)
\put(-50,-7){
\includegraphics[width=100mm]{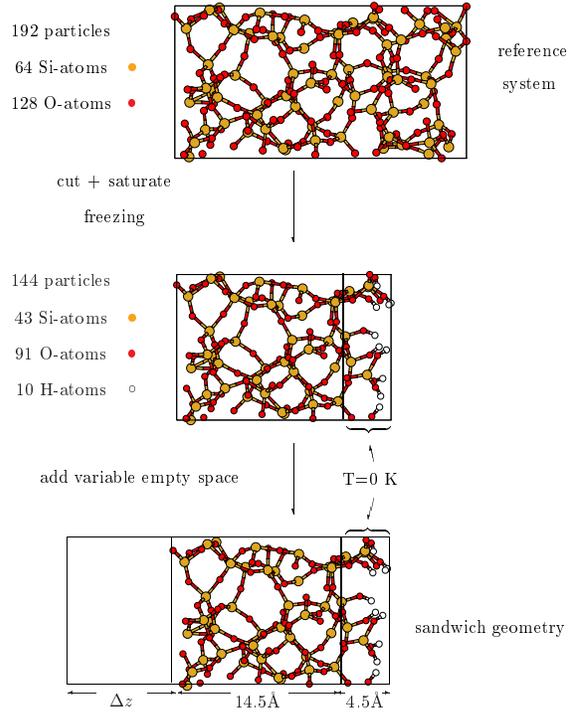}
}
\end{picture}
\caption{Procedure to create the used sandwich geometry}
\label{fig1}
\end{center}
\end{figure}
i) We start with a relatively large (bulk) silica system at
$T=3400$~K with PBC in three dimension (with box size $L_x=L_y=11.51$\AA\,
and $L_z'=23$\AA.  ii) We cut the system perpendicular to the
$z$-direction into two pieces. Without loss of generality we can assume
that the mean position of this cut is at $z=L_z'$. At this point it is
very important, that we cut only oxygen-silicon-bonds, such that we get
only free oxygen atoms at this interface (thus the interface will have
a bit of roughness). iii) These free oxygen atoms are now saturated
by hydrogen atoms. The place of these hydrogen atoms are chosen such
that each of the new oxygen-hydrogen bonds is in the same direction as
the oxygen-silicon bond which was cut and has a length of approximately
1 \AA. The interaction between the hydrogen atoms and the oxygen atoms
as well as the silicon atoms are described only by a Coloumbic term. The
value of the effective charge of the hydrogen atoms is set to $0.6$, which
ensures that the system is still (charge) neutral. iv) We make atoms which have a distance from this interface that is less than 4.5 \AA\
completely immobile, whereas the atoms that have a larger distance can
propagate subject to the force field. v) We add in $z$-direction an empty
space of $\Delta z=6.0$~\AA\, and thus generate a free surface at around
14.5~\AA. With this sandwich geometry we now can use periodic boundary
conditions in all three directions.  We have made sure that the value
of $\Delta z$ is sufficiently large that the results do not depend on it
anymore~\cite{mischler01}. Note that it is not advisable to choose $\Delta
z$ too large, since this would increase the cost of the CPMD simulation.
At the end of this procedure we have a system of 91 oxygen, 43 silicon
and 10 hydrogen atoms in a simulation box with L$_x$=L$_y$=11.51 \AA\/
and $L_z\approx 25$~\AA.

\section{CPMD-Simulation}
Since the time scale which is accessible to the CPMD-method is very
restricted, we have to combine the {\it ab initio} with classical
calculations. For this we first prepared a classical system as
described in the previous section, equilibrated it for about 1~ns which
is sufficient to equilibrate it completely~\cite{horbach99}. From
a subsequent production run with the same duration we picked 100
statistical independent configurations and used them to characterize the
static properties of the system with a high accuracy. Using a subset
of these configurations as starting points, we subsequently started
CPMD-simulation using the CPMD code developed in Stuttgart~\cite{cpmd}. For
the CPMD we used conventional pseudopotentials for silicon and oxygen and
the BLYP exchange-functions~\cite{trouiller91,lee88}. The electronic wave-functions
were expanded in a plane wave basis set with an energy cutoff of 60 Ry
and the equations of motion were integrated with a time step of 0.085 fs
for 0.2 ps. In the analysis of the CMPD data only those configurations
were taken into account that were produced later than 5 fs after the
start of the CPMD run in order to allow the system to equilibrate at
least locally~\cite{benoit00}.

In the analysis of the classical configurations we noted that typically
one of the three following situations is present on the surface:

\begin{itemize}
\item systems with no defects (i.e. all Si and O atoms are four and two-fold
coordinated, respectively)
\item systems with an undercoordinated oxygen atom and an undercoordinated silicon atom
\item systems with an overcoordinated oxygen atom and an undercoordinated oxygen atom
\end{itemize}

Therefore we picked for each case two BKS configurations and started
the CPMD runs.

\begin{figure}[ht]
\begin{center}
\unitlength1mm
\includegraphics[width=55mm]{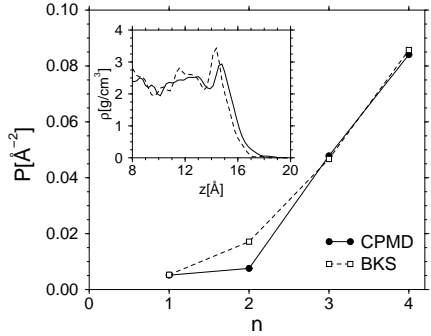}
\caption{Probability to find a ring of size $n$. Inset: $z-$dependence of the mass
density.}
\label{fig2}
\end{center}
\end{figure}

The largest differences between the results of the classical and of
the CPMD simulation are found for the short rings (n$<$5). (A ring
is a closed loop of $n$ consecutive Si-O segments~\cite{wright78}.) Figure~\ref{fig2}
shows the probability to find a ring of size $n$ for the case of BKS
and CPMD. We see that the BKS potential overestimates the frequency
with which a ring of size two occurrs by about a factor of two. Related
to this is the observation that the overshoot that is observed in the
$z-$dependent mass density profile is less pronounced in the case of
the CPMD than the one for the BKS (inset of Fig.~\ref{fig2}), since
two-membered rings are relatively dense.

From this inset we also see that the density profile for the CPMD extends
to larger $z$ values that the one for the BKS (by about 0.4~\AA). This
is because the BKS potential is not able to reproduce correctly the
density of silica at zero pressure.

\begin{figure}[ht]
\begin{center}
\unitlength1mm
\includegraphics[width=55mm]{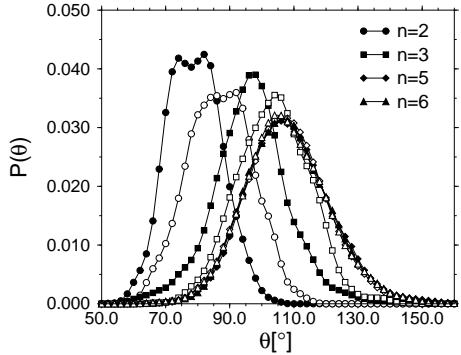}
\caption{Distribution of O-Si-O-angles for different ring sizes. BKS$=$filled symbols; CPMD$=$open symbols.}
\label{fig3}
\end{center}
\end{figure}

Another interesting result is the dependence of the distribution of angles
O-Si-O on the ring size (Fig.~\ref{fig3}). For large $n$, $n>$4, i.e. the
sizes which are normally found in the bulk \cite{rino93}, the results
of the two different methods are in good agreement \cite{benoit00}. For
smaller $n$, however, the mean O-Si-O-angle from CPMD is shifted to
larger values in comparison to the classical one. This shift becomes
more pronounced with decreasing $n$. Furthermore also the shape of the
distributions starts to become different if $n$ is small.

\begin{figure}[ht]
\begin{center}
\unitlength1mm
\includegraphics[width=55mm]{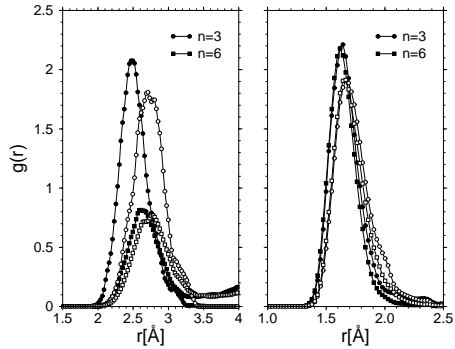}
\caption{Radial distribution function for different ring sizes. Left: O-O. Right Si-O. BKS$=$filled symbols; CPMD$=$open symbols.}
\label{fig4}
\end{center}
\end{figure}

This effect can be understood better by analyzing the partial radial
distribution functions $g(r)$ which are shown in Fig.\ref{fig4}. We
see that for the Si-O pair the curves from CPMD are shifted to larger
distances by about 0.04~\AA\, and that this shift is independent of
$n$. Also the $g(r)$ for the O-O pairs are shifted to larger $r$, but
this time the amount does depend on $n$. In particular we note that the
O-O distance is nearly independent of $n$ for the case of CPMD, whereas
it increases with $n$ for the case of BKS. These effects results in the
difference in the distribution of the O-Si-O angles if $n$ is small.

\section{Conclusion}

In this work we have investigated some structural properties of an
amorphous silica surface. In particular we have studied how these
properties depend on the simulation method: A classical simulation
with the potential proposed by van Beest {\it et al.} (BKS) and a
Car-Parrinello simulation (CPMD). We find that the structure on larger
length scales are independent of the method used, whereas the details of
structural elements on short scales (short rings, distribution function
for angles, etc.) differ. Thus this shows that it is probably necessary
to use {\it ab initio} methods if one wants to understand these systems
at short length scales in a quantitative way.

Acknowledgement: We gratefully acknowledge the financial support by the
SCHOTT Glaswerke Fond and the DFG under SFB 262 and the BMBF under grant N$^o$ 03N6015. We thank the NIC J\"ulich for a generous grant of computing time.

\end{document}